\documentstyle[pre,aps,multicol,epsf]{revtex}

\begin{document}
\draft

\title{Ground States of Two-Dimensional Polyampholytes}
\author{Eilon Brenner and Yacov Kantor}
\address{School of Physics and Astronomy, Tel Aviv University,
Tel Aviv 69 978, Israel}
\maketitle

\begin{abstract}
We perform an exact enumeration study of polymers formed from a
(quenched) random sequence of charged monomers $\pm q_0$, restricted
to a 2-dimensional square lattice. Monomers interact via a logarithmic (Coulomb) interaction. We study the ground state properties
of the polymers as a function of their excess charge $Q$ for all
possible charge sequences up to a polymer length $N=18$. We find
that the ground state of the neutral ensemble is compact and its energy
extensive and self-averaging. The addition of small excess charge causes an
expansion of the ground state with the monomer density depending only on $Q$. In an annealed ensemble the ground state
is fully stretched for any excess charge $Q>0$.
\end{abstract}



\begin{multicols}{2}
\narrowtext
\section{Introduction}
\label{intro}
Polymer systems have been subject to extensive study for several
decades \cite{FLO,PGD,GRO}. The interest in such
macromolecules arises from many sources, where one important reason is the natural occurrence of polymers in biological systems, for example
proteins. Random polymers are also good examples of disordered systems, sharing
common features with other physical systems exhibiting randomness or
frustration \cite{CHA,FRA}. An important class of polymers are {\it
  polyampholytes} (PA), where the chain is composed of positive and
negative charged monomers \cite{TAN}. Certain properties of proteins,
some of whose constituents, the amino acids, may be charged \cite{CRE},
may be understood by studying PAs.

Much attention has been devoted to studying the ground state
properties of polymers (or the {\it native state} of proteins) and
understanding which features determine their shape \cite{CHA,FRA}. In PAs, the long-range Coulomb interaction between the monomers may, in the
absence of screening, control the size and shape of the polymer,
dominating over other short-range interactions. For $d>4$, the behavior of the PA is
similar to that of random chains with short-range interactions, the
Coulomb interaction becoming an irrelevant perturbation \cite{KA3}. For $d<4$,
the effect of the Coulomb interaction is complicated. Higgs and Joanny
\cite{HIG}, and Wittemer {\it et al.\ }\cite{WIT}, expanding on
arguments previously made by Edwards {\it
  et al.\ }\cite{EDW}, pointed out, on the basis of the
Debye-H\"uckel theory \cite{MA}, that charge fluctuations along the
chain induce a net attractive interaction between segments of the
chain, leading to a collapse of the PA to
a compact shape. The resulting picture is of a {\it globule}
consisting of densely packed {\it blobs}. Each blob is a dilute region
of a weakly perturbed self-avoiding walk (SAW) of the size of the Debye-H\"uckel screening
length. Lowering the temperature, decreases the size of the
blobs. This behavior is not a phase transition but rather a gradual
increase of density of monomers with decreasing temperature. Charge correlations along the polymer may change the
nature of the collapse \cite{WIT}: a polymer with a quenched sequence
of alternating charges will undergo a $\theta$-transition \cite{VIC},
like polymers with short-range attraction interaction \cite{PG2}.

The application of the Debye-H\"uckel theory requires the strict
neutrality of the system \cite{MA}. If we assign to each monomer along
the chain a charge $q_i=\pm q_0$, then an excess charge $Q=\sum_{i=1}^Nq_i$ may be
defined. A random PA will not necessarily be neutral but may carry an
excess charge $Q$, typically on the order of $q_0\sqrt N$. Gutin and Shakhnovich \cite{GU2}, and Dobrynin and Rubinstein
\cite{DOB} treated the problem of a charged PA. They found an
expansion of the ground state due to excess charge and describe it in
the terms of an elongated globule with a $Q$-dependent aspect ratio.
Observing that for neutral quenches ($Q=0$) the ground state is
compact, Kantor and Kardar \cite{KA5,KA6} described the ground state
properties of a charged PA in an analogy to a drop of a
fluid with a charge distributed in it: The energy of a neutral drop may be described by an extensive form with a
surface tension accounting for the compact spherical
shape. The excess charge $Q$ is the important parameter
determining the properties of the drop (or compact PA). A
phenomenological description of the energy is given by 
\begin{equation}
\label{phenom}
E=-\epsilon_cV+\gamma S+Q^2F(R)\ ,
\end{equation}
where the first term is the extensive contribution to the energy
(proportional to the volume $V$), the second term is a correction
proportional to the surface area $S$, while the third term is the Coulomb interaction,
where $F(R)$ represents the dependence of the Coulomb potential on the
linear size of the drop $R$ and is given (up to dimensionless
prefactors of order unity) by:
\begin{equation}
F(R)=\left\{
\begin{array}{ll}
\frac{1}{R^{d-2}}& \mbox{for $d\neq2$,}\\
-\ln R&\mbox{for $d=2$.}
\end{array}\right.
\end{equation}
Increasing the charge
$Q$ creates an outward pressure that seeks to expand the drop,
competing with the surface tension. At a certain excess charge, called
the Rayleigh charge $Q_R$, the pressure difference between the inside
and outside of the drop vanishes and the drop becomes locally
unstable to elongation. In a $d$-dimensional PA this happens at $Q_R^2\approx
q_0^2N^{2-3/d}$ \cite{KA6}. However,
even at a lower excess charge it might be energetically
favorable to split a drop into two distant droplets each carrying half the
excess charge leading to a global instability. The critical charge
$Q_c$ when this occurs is found to be $Q_c^2=\alpha Q_R^2$, where the
proportionality constant depends on $d$. This is
the excess charge above which the PA should expand. In 3D, both the
typical excess charge and the critical charge scale as $\sqrt N$,
hence, a typical PA will be stretched. Numerical studies using
Monte-Carlo methods \cite{KA5} and exact enumeration techniques
\cite{KA6} support these qualitative predictions in 3D. Once the critical charge is
exceeded, the PA expands, but it does not have the freedom of the drop
to disintegrate, since the polymer is a connected object. It can,
however, be approximately described as a {\it necklace} \cite{KA5} of
weakly charged globules connected by highly charged
strands of a PA.

While the 3D case is important from the practical point of view,
certain aspects cause conceptual difficulties (and consequently
difficulties in the interpretation of Monte-Carlo results): In 3D,
both $Q_R$ and the critical charge which determines the
high-temperature behavior of a PA \cite{KA3} and the typical fluctuation of the excess charge
$Q$ between the quenches, have the same scaling -- all three
quantities are proportional to $\sqrt N$. It is, therefore, beneficial
to study PAs in other space dimensions $d$. We study a model of a PA,
a 2-dimensional (2D) randomly charged SAW on a 2D square lattice. Although
this model significantly simplifies the complexity of real physical
systems, we hope to gain some physical insight and understand the
relevant mechanisms governing the behavior of the system. In Sec.\
\ref{model} we elaborate on the specific model we study.

The logarithmic behavior of a 2D Coulomb potential creates
difficulties which are absent in $d>2$: It is even not self-evident
that the low temperature configurations have an extensive energy and
surface tension, i.e.\ that the description of Eq.\ (\ref{phenom}) is
applicable in 2D. In Sec.\ \ref{neut} we will show that the extensive description of the energy
applies to our 2D model for $Q=0$. In Sections
\ref{excess},\ref{annealed} we study the effect of excess charge on 2D
PAs. Although the short chains studied prevent the verification of the
necklace description, we see that any deviation from neutrality leads
to an expansion of the ground state. We find that the system is much more
susceptible to excess charge in 2D than in 3D.

\section{The Model and Method}
\label{model}
We model the polymer as a SAW on a 2D square
lattice of spacing $a$. Monomer $i$ along the chain has a charge
$q_i$ which is randomly assigned a value $\pm q_0$. The charges
interact via a 2D Coulomb potential
\begin{equation}
\label{ham}
E=-\sum_{\left <ij\right >}q_iq_j\ln \frac{r_{ij}}{r_0}\ ,
\end{equation}
where $r_{ij}$ is the distance between the $i$th and $j$th monomers and $r_0$ is an
arbitrary constant setting the reference point of the energy. The study
of a 2D system allows us to extend our study to longer polymers than
was possible in 3D \cite{KA6}. In addition, for short
2D chains, the surface to volume ratio, responsible for many of the
properties of the polymer, is similar to that of long, realistic 3D
chains \cite{CH2}. Everywhere (except Sec.\ \ref{annealed}), we
consider a fixed charge sequence, i.e.\ quenched disorder. Note that
the energy (\ref{ham}) is symmetric under charge
conjugation, so we study only overall neutral and positively charged quenches.

In this study we apply the complete enumeration method: We
create all possible spatial configurations of the chain and calculate
exact quantities for each configuration. This allows us to find the ground
state of the system and, when studying higher temperatures (results we
do not report here), to compute exact thermodynamic averages. While
the number of SAWs grows exponentially with $N$ \cite{PGD} limiting us to relatively short chains, this is
the only method for studying exact ground state properties, since
Monte Carlo methods fail
to equilibrate at low temperatures. Up to $N=18$ we enumerated all possible charge
sequences. For a PA 18 monomers long, after taking into account basic
symmetries, there are 5,808,335 spatial configurations for each of the
77,819 possible quenches (unrelated by symmetry). Enumerating all
configurations for all the quenches of $N=18$ required approximately 20 days of CPU time on a
Silicon Graphics R10000 workstation. Up to a length of 26 monomers
we enumerated only partial samples of all possible quenches (1000,
500, 100 and 50 random neutral quenches for lengths $N=$ 20, 22, 24 and
26 respectively). A 26
monomer long chain has 15,435,169,364 possible spatial configurations,
unrelated by symmetry. Similar model and method were employed by Kantor and
Kardar \cite{KA6} to study 3D systems up to $N=13$.

There is an important point to note regarding the energy of a 2D
system with a fixed number of charges. We may express the
distances as $r_{ij}=ar_{ij}'$, where $r_{ij}'$ is a dimensionless
distance (in lattice constants), and re-write Eq.\ (\ref{ham}) in the form:
\begin{equation}
\label{eng}
E=-\sum_{\left <ij\right >}q_iq_j\ln
r_{ij}'-\frac{1}{2}\left(Q^2-q_0^2N\right)\ln\frac{a}{r_0}\ .
\end{equation}
The first term depends on the specific spatial conformation. The second term is a constant reference point, independent of
the spatial configuration or charge sequence along the polymer, and is a
function of the excess charge only. Here on, we take
$r_0=a$, so that the second term vanishes and calculate only the first term
as the ``energy.'' Within an ensemble of the same excess charge, this
has no effect since we are interested only in
energy differences. However, it must be kept in mind that for
different excess charges and lengths, the choice of $r_0$ may set different reference points, so energies cannot be compared.

\section{Ground State Spectrum of Neutral Polyampholytes}
\label{neut}
Following the reasoning given in Sec.\ \ref{intro} and empirical
observations we expect the ground state of neutral PAs to be
compact. It would be natural to consider an extensive form of the
energy of the ground state with a surface correction. Phenomenologically, this can
described as:
\begin{equation}
\label{engeq}
\overline{E}=A_1q_0^2N+A_2q_0^2\sqrt N\ ,
\end{equation}
where $\overline{\cdots}$ represents an average over
quenches. The
second term in (\ref{engeq}) represents the 2D surface correction. Unlike short-range
interacting systems, it is not obvious that the 2D Coulomb
interaction produces such a form of $\overline E$. For instance, the
energy of a {\em randomly} selected spatial configuration of a
neutral sequence is expected to be of order $N\ln N$, i.e.\ it grows
faster than $N$.

\begin{figure}[]
\epsfysize=16\baselineskip
\centerline{\hbox{
      \epsffile{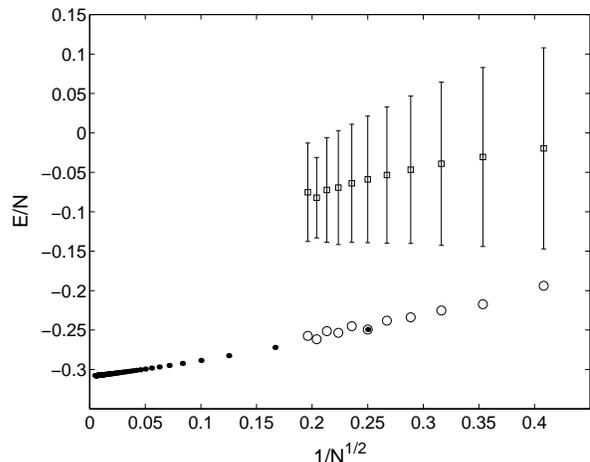}  }}
\caption [A Picture]
         {\protect\footnotesize Ground state energies per monomer
           \cite{FIG} for
           neutral sequences of lengths 6-26 monomers. For each
           length, boxes represent the mean value over
           all quenches studied, while error bars represent the standard 
           deviation. The results for the alternating sign
           quench are represented by open circles. The energy of a ``salt crystal'' plane is represented by dots.}
\label{inter}
\end{figure}

A convenient comparison point for the ground state energies is provided
by a checkerboard configuration (``salt crystal'') of charges, which
is the ground state of a 2D Coulomb gas on a lattice \cite{LEE}. It
can be shown that for such an arrangement of charges the energy is
extensive. Dots
in Fig.\ \ref{inter} depict the energy per monomer for finite
``salt crystal'' configurations forming a square consisting of $N=4n^2$ ($n=$2, 3, 4,
...) charges. The energy of such a crystal is given by
\begin{equation}
E=-0.31q_0^2N+0.26q_0^2\sqrt N\ .
\end{equation}
Not surprisingly, in PAs the ground state of quenches which are an
alternating sequence of positive and negative charges (open circles in
Fig.\ \ref{inter}), follow the same law. However a typical quench cannot assume a salt-crystal-like ground state. Results of evaluating all ground state energies for the
different quenches are depicted in Fig.\ \ref{inter}. For each length,
we display the mean and standard deviation of the ground state energy
taken over the different quenches. The mean energies may be fitted to an
extensive form: 
\begin{equation}
\overline{E}=(-0.133\pm0.005)q_0^2N+(0.30\pm0.02)q_0^2\sqrt N\ .
\end{equation}
Results must be considered with caution due to the rather short chains
studied. A definite exclusion of possible logarithmic corrections
would require the consideration of chain lengths larger by at least
one order of magnitude. Nevertheless, since the energy of a salt
crystal bounds from below the ground state energies, we believe that the extensive
form is valid.
\begin{figure}[]
\epsfysize=16\baselineskip
\centerline{\hbox{
      \epsffile{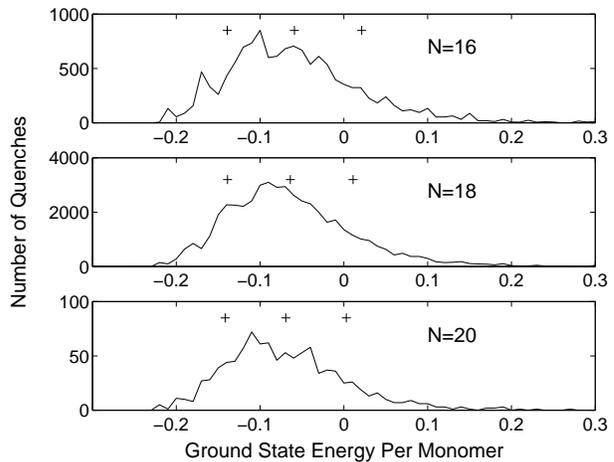}  }}
\caption [A Picture]
         {\protect\footnotesize A histogram of the ground state energy
           per monomer \cite{FIG} for all quenches of lengths 16 and 18 monomers
           and a partial sample of quenches of length 20 monomers. Crosses
           indicate mean and standard deviation limits. Bin size is $0.01q_0^2$.}
\label{hist}
\end{figure}

\begin{figure}[]
\epsfysize=16\baselineskip
\centerline{\hbox{
      \epsffile{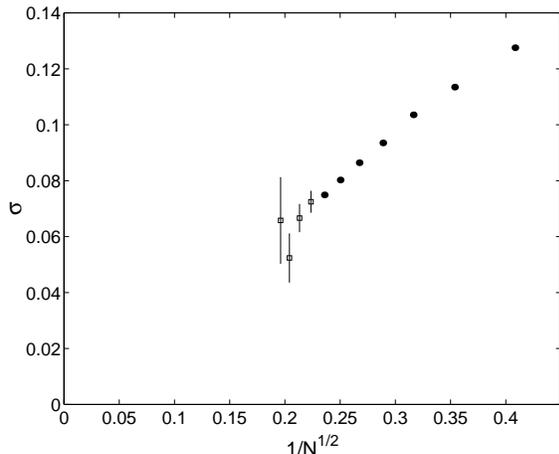}  }}
\caption [A Picture]
         {\protect\footnotesize Standard deviation of the ground state
           energy per monomer $\sigma$ \cite{FIG}, as a function of $1/\sqrt N$, for
           neutral quenches of lengths 6, 8,...,26
           monomers. Error bars denote 98\% confidence levels when the
           standard deviation was estimated for partial samples, in
           the four leftmost cases.}
\label{self}
\end{figure}

The distribution of the ground state energies per monomer, for the
ensemble of quenches, is relatively
broad even for the largest $N$s considered: Fig.\ \ref{hist} depicts
the distribution of the ground state energies of $N=$16 and 18 (all
possible quenches), and 20 monomers (a partial sample only). Nevertheless, the standard deviation $\sigma$ of the energy per
monomer gradually decreases with increasing $N$. In systems with
short-range interactions the energy of a large random system may
deviate from the ensemble-averaged mean by an amount $\sim\sqrt
N$. This is, usually, a consequence of the fact that the total
energy of the system is (approximately) a sum of the energies of the
subsystems: E.g., the total energy is a sum of energies obtained by
dividing the system into two halves. This property is referred to as
self-averaging. Fig.\ \ref{self} demonstrates the linear dependence of
the standard deviation of the energy per monomer $\sigma$ on $1/\sqrt
N$, which means that the fluctuations of the total energy are
$\sim\sqrt N$. Thus, we may treat our system as
self-averaging. However, if we split a large 2D {\it neutral} PA into
two equal parts, usually none of the halves will be neutral. On the contrary,
typically each half will be charged by an equal and opposite charge
$Q'\sim\sqrt N$, and, thus, have electrostatic energy of order
$-Q'^2\ln R'\sim -N\ln N$. Moreover, as will be shown in the next
section, such charged sub-chains will be expanded, i.e.\ the
spatial conformations of the sub-chains will not resemble the
configuration of the entire PA. Nevertheless, we may use a slightly
modified concept: In a very long neutral PA, one can, usually, find several
special points, such that splitting the the PA at those points will
divide it into several large {\it neutral} sub-systems (see, e.g.,
Ref.\ \cite{FIK}). The total energy of a PA can now be regarded as the sum
of the energies of these separate segments.

\section{Excess Charge Effects}
\label{excess}
\begin{figure}[]
\epsfysize=16\baselineskip
\centerline{\hbox{
      \epsffile{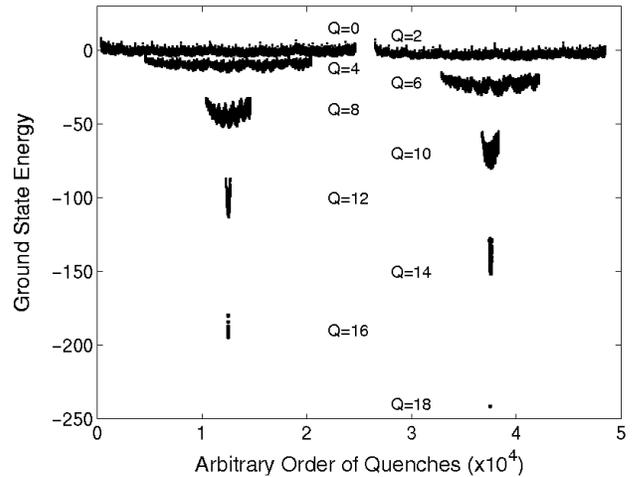}  }}
\caption [A Picture]
         {\protect\footnotesize Ground state energies for all quenches
           of length 18 monomers \cite{FIG}. Abscissa depicts an arbitrary
           order of quenches. Annotations denote the excess charge of
           the bands the energies group up into.}
\label{bands}
\end{figure}

Non-vanishing excess charge $Q$ has a profound effect on the energies
and spatial conformations of PAs. Fig.\ \ref{bands} depicts the ground
state energies of all quenches of PAs of $N=18$. Every point in this
figure corresponds to a different quench. The ground state energies
split into bands, each of which corresponds to a different $Q$. The
distance between the band of neutral quenches and the rest of the
bands increases approximately as $Q^2$ due to the appearance of a
long-range electrostatic term $-Q^2\ln R$, where the length $R$
characterizes the spatial extent of the PA in the ground state. (The
length $R$ depends on $Q$, however, $\ln R$ has a minor effect on the
leading $Q^2$ dependence.) One should keep in mind that, in general,
the absolute values of the energy difference between the bands depends
on the choice of the reference point $r_0$ in Eq.\ (\ref{eng}). We
note that bands remain relatively narrow, i.e.\ the ground state energy
weakly depends on the details of the specific quench. Only when $Q$
approaches $q_0N$, the bands broaden, i.e.\ the details of the quench
have a more significant influence on the overall energy.
\begin{figure}[]
\epsfysize=16\baselineskip
\centerline{\hbox{
      \epsffile{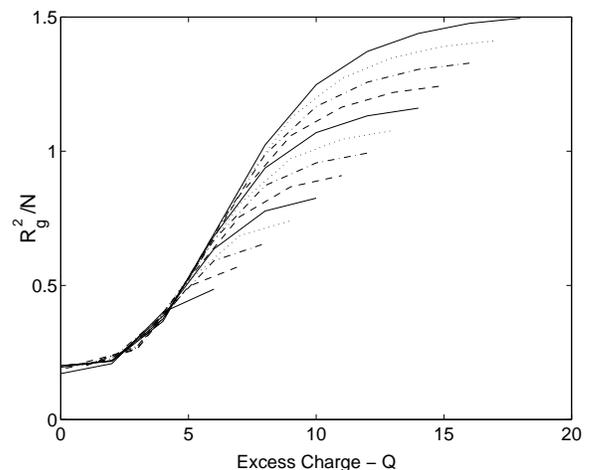}  }}
\caption [A Picture]
         {\protect\footnotesize $R_g^2/N$ {\it vs.\ }$Q$ \cite{FIG}. Curves
           represent different lengths ($N=$6, 7,...,17,18 from bottom
           to top). Note the collapse of the curves at low $Q$ and
           gradual deviation as $Q$ is increased.}
\label{crit}
\end{figure} 

Excess charge is an important parameter determining the shape of the
ground state as well. Specifically, it determines whether the ground
state is compact or expanded. We measure the typical size of a
configuration by its root-mean-square size called the radius of gyration $R_g$. A compact shape is defined by
$R_g\sim N^{1/d}$. In 2D this means that $R_g^2/N$ is independent of $N$. In Fig.\
\ref{crit} we plot $R_g^2/N$ as a function of $Q$ for the different
values of $N$. Each point is an average over all possible quenches of
the specific $N$ and $Q$. The most striking feature of Fig.\ \ref{crit} is that for small $Q$ all the curves,
for different values of $N$, collapse onto the same curve. Thus, even
for very large $N$ a minute charge $Q$ can significantly increase $R_g$. This means that among the
ensemble of random quenches in the $N\rightarrow\infty$ limit a
vanishing portion remains compact. Even if the ensemble is biased towards
neutrality, i.e. $Q_{\rm typical}\sim N^x$, where $1\gg x>0$, the resulting
configurations will be expanded. Only exceptional quenches, where the excess charge is very small
even for large $N$, will remain
compact. This result may also be incorporated
within the charged drop model for a PA. In 2D $Q_R\sim N^{1/4}$, but the logarithmic potential creates a certain
difference with respect to higher dimensions: Because the potential
has no finite asymptotic value, the charged drop will always find it favorable
to disintegrate into distant droplets with an infinite energy
gain. Hence, in a continuum model $Q_c=0$, and we except an expansion
in the size of the ground state of the PA for any excess charge. This
is very different from 3D systems \cite{KA5,KA6}, where
$Q_c\propto Q_R\sim\sqrt N$ so a finite portion of all random quenches
remains compact. Once the critical charge is exceeded, the ground
state will begin expanding but does not immediately become fully stretched. Numerical results
show that in 2D for
$Q\sim N^{0.7}$ the resulting configuration is completely stretched,
i.e.\ $R_g\sim N$. This implies that if the charge is a finite fraction
of the size, i.e. $Q\sim N$, the ground state is fully stretched.

\section{The Annealed Ensemble}
\label{annealed}
\begin{figure}[]
\epsfysize=15\baselineskip
\centerline{\hbox{
      \epsffile{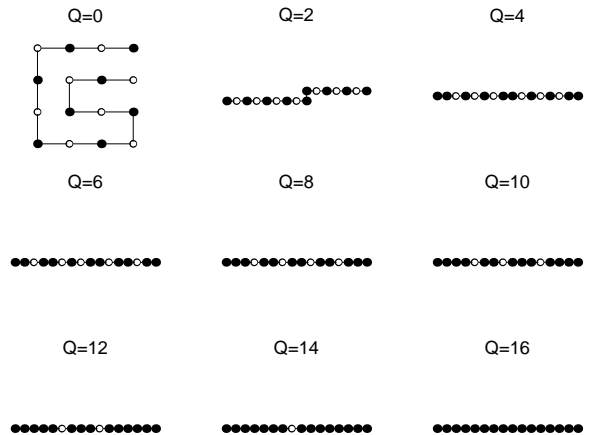}  }}
\caption [A Picture]
         {\protect\footnotesize Ground state configuration for
           annealed sequences of length 16, with an overall excess
           charge restriction $Q$. Opposite charge types are denoted
           by open and full circles.}
\label{ann}
\end{figure}

Thus far we have considered quenched sequences in which charges are
fixed in position along the chain. We now remove this restriction,
allowing charges to exchange positions along the chain, maintaining the overall excess charge and
number. (Two charges cannot be located at the same site, thus creating
a double or vanishing charge.) We shall refer to this as an {\it annealed} ensemble of
PAs. The ground state configurations for the ensembles of $N=16$ and
different values of Q are shown in Fig.\ \ref{ann}.

For the neutral ensemble ($Q=0$) the ground state is naturally a
sequence of charges alternating in sign that form a ``salt crystal''
configuration, in the same way that free charges form such a crystal \cite{LEE}. This spatial
arrangement of the charges can be obtained by many different
spatial configurations of the chain, making the ground state highly
degenerate. For any non-zero excess charge we observe a sharp
transition and the ground state has a fully stretched rod-like
shape. For the sizes considered here, the ground state configuration
and charge sequences were unique.

Previously, we have seen that for a finite $Q$, the ground state
configuration of the quenched ensemble expands. In the annealed case
the expansion is much more abrupt: The PA is completely stretched for
any minute $Q$. It was suggested \cite{KA5} that an annealed PA in 3D, with a large enough excess
charge, might find it favorable to expel the excess charge in the form
of highly charged fingers, while most of the monomers remain in a
globule. In 2D the behavior seems to be quite different: Small excess
charge suffices to stretch the PA into a straight line. The positions
of the charges on that line are determined as follows: in general the
charges attempt to maintain an alternating sequence; this not strictly
possible because of nonvanishing $Q$. The excess charge is spread out
almost uniformly (up to logarithmic corrections) along the PA.

These results should be regarded with caution. Contrary to the
quenched ensemble, where the ground state is averaged over many
quenches, thus smoothing out lattice effects, the ground state of the
annealed ensemble corresponds to a single configuration of a single
sequence. The chains studied might not be long enough to overcome this
problem.

\section{Discussion}
We studied the ground state properties of a model 2D PA. We found that neutral quenches have a compact, dense ground state, whose
energy can be described as an extensive energy term with a surface
correction. The ground state energies of the different quenches may be
described by energy bands, each band corresponding to a certain excess charge. The specific sequence of
charges is of a small influence, creating small fluctuations about the
mean within the
bands (for neutral quenches the ground state energy is even
self-averaging over all possible quenches). When the charge sequence
is annealed (not quenched) the shape of the ground state is much more sensitive to excess charge, i.e.\ the PA is fully stretched for any
$Q\neq0$. The results are different than those obtained for a
similar 3D model \cite{KA6}. The nature of the 2D Coulomb interaction accounts for this difference. Both cases, however, match the
expectations derived from an analogy with a charged drop. The main
drawback of the method we applied throughout this study is the limited
size of the systems we can fully investigate. Although we surpassed
the maximal length studied in 3D, some results may be
suspected as arising from finite size and lattice effects. A major
increase in computer capabilities is required to carry this method to
larger systems.

Several authors \cite{VIC,LV1} suggested that, at low temperatures, a PA
may undergo a freezing transition as do some models of polymers with a
short-range interaction \cite{FRA,BRY,DIN} and other disordered
systems. Should this be the case, then the ground state properties do
not provide the full description of the zero-temperature behavior of
the system. Our investigation of such a transition was not conclusive,
but indicates its absence, at least in the sense
that the requirements of the random energy model \cite{DER} are
violated. This is in agreement with predictions of Pande {\it et al.\ }\cite{PA1}
that long-range interacting systems would not undergo a glass
transition.

\section*{Acknowledgments}
EB would like to thank I.\ Golding for useful
discussions. This work was supported by the US-Israel BSF under grant No.\ 96-46.


\end{multicols}

\begin{references}
\bibitem{FIG}
In all figures in this paper lengths are measured in units of $a$,
charges in units of $q_0$ and energies in units of $q_0^2$.
\end{references}
\end{document}